# Chemical pressure effect on superconductivity of $BiS_2$-based $Ce_{1-x}Nd_xO_{1-y}F_yBiS_2$ and $Nd_{1-z}Sm_zO_{1-y}F_yBiS_2$

Joe Kajitani*, Takafumi Hiroi, Atsushi Omachi, Osuke Miura, and Yoshikazu Mizuguchi

Department of Electrical and Electronic Engineering, Tokyo Metropolitan University, Hachioji, 192-0397

Corresponding author: Joe Kajitani (kajitani-joe@ed.tmu.ac.jp)

**Abstract**

We have systematically investigated the crystal structure and the magnetic properties of $BiS_2$-based superconductor $Ce_{1-x}Nd_xO_{1-y}F_yBiS_2$ ($x$ = 0 - 1.0, $y$ = 0.3, 0.5 and 0.7) and $Nd_{1-z}Sm_zO_{1-y}F_yBiS_2$ ($x$ = 0 - 0.8, $y$ = 0.3, 0.5 and 0.7). In the $REOBiS_2$ system, both crystal structure and physical properties are tunable by substituting with the different RE (RE = rare earth) elements, having a different ionic radius, such as Ce, Nd and Sm.

In the $Ce_{1-x}Nd_xO_{1-y}F_yBiS_2$ system, bulk superconductivity is observed for $x$ = 1.0 with $y$ = 0.3 and $x \geq 0.5$ with $y$ = 0.5. The transition temperature ($T_c$) increases with increasing Nd concentration. The highest $T_c$ is 4.8 K for $x$ = 1.0 with $y$ = 0.5 in the $Ce_{1-x}Nd_xO_{1-y}F_yBiS_2$ system. By the Nd substitution for Ce, lattice contraction along the $a$ axis is generated while the $c$ axis does not show a remarkable dependence on Nd concentration. The lattice constant of $c$ decreases with increasing F concentration.

Furthermore, we found that the Nd site can be replaced by smaller Sm ions up to $z$ = 0.8 in $Nd_{1-z}Sm_zO_{1-y}F_yBiS_2$. Bulk superconductivity is observed within $z \leq 0.8$ for $y$ = 0.3 and $z \leq 0.6$ for $y$ = 0.5. The $T_c$ increases with increasing Sm concentration. The highest $T_c$ is 5.6 K for $z$ = 0.8 with $y$ = 0.3. With increasing Sm concentration, the lattice constant of $a$ decreases while the lattice constants of $c$ does not show a remarkable dependence on Sm concentration.

We found that the chemical pressure generated by systematic solution of the RE site in the blocking layer commonly induces lattice contraction along the $a$ axis in $Ce_{1-x}Nd_xO_{1-y}F_yBiS_2$ and $Nd_{1-z}Sm_zO_{1-y}F_yBiS_2$. The obtained results indicate that both optimal F concentration and uniaxial lattice contraction along the $a$ axis generated by chemical pressure are essential for the inducement of bulk superconductivity in the $REO_{1-y}F_yBiS_2$ system.

## 1. Introduction

Layered materials have been studied in the field of superconductivity because superconductors with a high transition temperature ($T_c$) had been discovered in layered materials, as were the Cu-oxide[1] and the Fe-based superconductors[2]. Recently, several types of $BiS_2$-based layered superconductors such as $Bi_4O_4S_3$[3], $REO_{1-x}F_xBiS_2$ (RE = La, Ce, Pr, Nd)[4-9] and $Sr_{1-x}La_xFBiS_2$[10,11] were discovered. The crystal structure of the $BiS_2$-based superconductor is composed of an alternate stacking of double $BiS_2$ conduction layers and blocking layers such as $Bi_4O_4(SO_4)_{1-x}$, $RE_2O_2$, or $Sr_2F_2$ layers. Band calculation suggested that the parent material is a band insulator, and the $BiS_2$-based materials become superconductive when electron carriers were generated at the Bi-6p orbitals within the $BiS_2$ conduction layers[3,12]. $LaO_{1-x}F_xBiS_2$ is a typical $BiS_2$-based superconductor. The parent phase $LaOBiS_2$ is semiconducting, and partial substitution of the O site by F generates electron carrier within the $BiS_2$ layers. Recently, it was reported that optimization of crystal structure is important for the inducement of superconductivity in the $LaO_{1-x}F_xBiS_2$ system as well as electron carriers. Namely, the crystal structure and the superconducting properties strongly correlate in the $BiS_2$ family. One of the notable characteristics is positive pressure effect on $T_c$ [13-18]. The $T_c$ of $LaO_{0.5}F_{0.5}BiS_2$ largely increases from 3 K to 11 K by application of external high pressure or sample synthesis under high pressure.

On the other hand, Substitution with isovalent elements generates chemical pressure. It is useful for discussing the relationship between crystal structure and superconducting properties to study the chemical pressure effects on superconductivity in this system because superconducting properties are tunable by application of chemical pressure in several layered materials. Recently, we have studied the chemical pressure effect on superconducting properties in $BiS_2$-based $Ce_{1-x}Nd_xO_{0.5}F_{0.5}BiS_2$[20]. As a result of the study, we found that the lattice contracts along the *a* axis, and bulk superconductivity was induced with increasing Nd concentration ($x$). To clarify the relationship between crystal structure and superconducting properties in detail and to increase $T_c$, we have systematically investigated chemical pressure effect in $Ce_{1-x}Nd_xO_{1-y}F_yBiS_2$ and $Nd_{1-z}Sm_zO_{1-y}F_yBiS_2$ systems.

## 2. Experimental details

The polycrystalline samples of $Ce_{1-x}Nd_xO_{1-y}F_yBiS_2$ were prepared by solid-state reaction using powders of $Ce_2S_3$ (99.9 %), $Nd_2S_3$ (99 %), $Bi_2O_3$ (99.99 %), $BiF_3$ (99.9 %), $Bi_2S_3$ and grains of Bi (99.999 %). The polycrystalline samples of $Nd_{1-z}Sm_zO_{1-y}F_yBiS_2$ were prepared by solid-state reaction using powders of $Nd_2S_3$ (99 %), $Sm_2S_3$ (99.9 %), $Bi_2O_3$ (99.99 %), $BiF_3$ (99.9 %), $Bi_2S_3$ and grains of Bi (99.999 %). The $Bi_2S_3$ powder was prepared by reacting Bi (99.999 %) and S (99.99 %) grains in an evacuated quartz tube. Other chemicals were purchased from Kojundo-Kagaku laboratory. The starting materials with a nominal composition of $Ce_{1-x}Nd_xO_{1-y}F_yBiS_2$ and $Nd_{1-z}Sm_zO_{1-y}F_yBiS_2$ were well-mixed, pressed into pellets, sealed in an evacuated quartz tube, and heated at 700ºC for 10h. The obtained products were ground, sealed in an evacuated quartz tube and heated again with the same heating conditions for crystal homogenization. The obtained samples were characterized by X-ray diffraction using the $\theta$–$2\theta$ method. The temperature dependence of magnetization was measured by a superconducting quantum interface device (SQUID) magnetometer with an applied magnetic field of 5 Oe after both zero-field cooling (ZFC) and field cooling (FC).

## 3. $Ce_{1-x}Nd_xO_{1-y}F_yBiS_2$

Figure 1(a), (c) and (e) show X-ray diffraction patterns for $Ce_{1-x}Nd_xO_{1-y}F_yBiS_2$ with $y$ = 0.7, 0.5 and 0.3, respectively. Almost all of the observed peaks are explained using the tetragonal *P*4/*nmm* space group. The numbers displayed with the X-ray patterns indicate Miller indices. To analyze the change in the lattice constant of *a* axis, the enlarged X-ray diffraction patterns near the (200) peak for the $Ce_{1-x}Nd_xO_{1-y}F_yBiS_2$ samples with $y$ = 0.7, 0.5 and 0.3 are shown in Fig. 1(b), (d) and (f), respectively. It is clear that the peak position of the (200) peak systematically shifts to higher angles with increasing Nd concentration ($x$) for $y$ = 0.7, 0.5 and 0.3. These results indicate the lattice constant of *a* decreases with increasing Nd concentration ($x$). In contrast, the (00*l*) peaks do not show a remarkable dependence on Nd concentration, indicating that the substitution of the RE site does not affect the *c* axis length. To clarify the relationship between RE (RE = Ce and Nd) concentration and lattice constants of *a* and *c*, the lattice constants were calculated using the observed peak positions of the (200) and (004) peaks. The calculated lattice constants *a* and *c* for all the samples of $Ce_{1-x}Nd_xO_{1-y}F_yBiS_2$ ($y$ = 0.3, 0.5 and 0.7) are plotted in Fig. 2(a) and (b). With increasing Nd concentration ($x$), the lattice constant of *a* monotonously decreases for all F concentrations ($y$) while the lattice constant of *c* does not show a remarkable dependence on Nd concentration. However, the lattice constants of *c* decrease with increasing F concentration. This result indicates the partial substitution of O by F generates lattice contraction along the *c* axis while the partial substitution for Ce by Nd generates lattice contraction along the *a* axis.

Figure 3(a) and (b) show the temperature dependence of magnetic susceptibility for $Ce_{1-x}Nd_xO_{0.3}F_{0.7}BiS_2$. The onset of $T_c$ ($T_c^{onset}$) was defined as a temperature where the temperature derivative of magnetic susceptibility ($d\chi/dT$) becomes negative in ZFC condition. The $\Delta\chi$ was defined as an absolute value of the difference between $\chi$ at $T_c^{onset}$ and $\chi$ at 2 K to investigate the bulk characteristics of the samples. In $Ce_{1-x}Nd_xO_{0.3}F_{0.7}BiS_2$, a superconducting transition is observed within a wide Nd-concentration range of $x$ = 0.0 - 1.0 as shown in Fig. 3(a) and (b) though the diamagnetic signals are weak: the value of $\Delta\chi$ is lower than 0.05 emu/Oe·cm$^3$. The $T_c^{onset}$ increases with increasing Nd concentration and reaches 4.8 K for $x$ = 1.0 ($NdO_{0.3}F_{0.7}BiS_2$). Furthermore, for $x$ = 0 - 0.8, the magnetic transition with a magnetic transition temperature ($T^{mag}$) of about 7.5 K is observed. The $T^{mag}$ was difined as the temperature at which $\chi$ abruptly increases along with the appearance of magnetic ordering of the RE site. The results are similar with the previous studies on magnetic properties of $CeO_{1-x}F_xBiS_2$[5,6,21]. In fact, the magnetism is due to the ordering of the Ce moment of the blocking layer. It is quite interesting that even the sample with $x$ = 0.8, which contains 80 % of Nd ions, shows magnetic transition like ferromagnetism observed in $CeO_{1-x}F_xBiS_2$.

In $Ce_{1-x}Nd_xO_{0.5}F_{0.5}BiS_2$, bulk superconductivity is observed within $x \geq 0.5$ as shown in Fig. 3(c). The largest value of $\Delta\chi$ is observed for $x$ = 1.0. The $T_c^{onset}$ increases with increasing Nd concentration and reaches 4.8 K for $x$ = 1.0. The magnetism observed in $Ce_{1-x}Nd_xO_{0.3}F_{0.7}BiS_2$ does not appear in $Ce_{1-x}Nd_xO_{0.5}F_{0.5}BiS_2$.

In $Ce_{1-x}Nd_xO_{0.7}F_{0.3}BiS_2$, filamentary superconductivity is observed for $x$ = 0.8, and bulk superconductivity is observed only for $x$ = 1.0 with $T_c^{onset}$ = 3.6 K as shown in Fig. 3(d). The magnetism observed in $Ce_{1-x}Nd_xO_{0.3}F_{0.7}BiS_2$ does not appear in $Ce_{1-x}Nd_xO_{0.7}F_{0.3}BiS_2$.

Figure 4 shows Nd concentration dependence of $T_c^{onset}$ for $Ce_{1-x}Nd_xO_{1-y}F_yBiS_2$ ($y$ = 0.3, 0.5 and 0.7). The symbols of × indicate that the sample at this point does not exhibit a superconducting transition at above 2 K. With increasing Nd concentration, the $T_c^{onset}$ increases for all system, and the maximum value of $T_c^{onset}$ is 4.8 K for $x$ = 1.0 with $y$ = 0.5 in $Ce_{1-x}Nd_xO_{1-y}F_yBiS_2$. Furthermore, the Nd concentration, at which superconductivity is induced, depends on the F concentration ($y$). For $Ce_{1-x}Nd_xO_{0.3}F_{0.7}BiS_2$, superconducting transition is observed for $x$ = 0 – 1.0 but the induced superconductivity is not bulk in nature. For $Ce_{1-x}Nd_xO_{0.5}F_{0.5}BiS_2$, bulk superconductivity is induced for $x \geq 0.5$. For $Ce_{1-x}Nd_xO_{0.7}F_{0.3}BiS_2$, bulk superconductivity is observed only for $x$ = 1.0. In the $Ce_{1-x}Nd_xO_{1-y}F_yBiS_2$ system, a higher Nd concentration ($x$) and optimal F concentration ($y$) are important for inducement of bulk superconductivity.

**4. $Nd_{1-z}Sm_zO_{1-y}F_yBiS_2$**

To further investigate the effects of chemical pressure, we prepared polycrystalline samples of $Nd_{1-z}Sm_zO_{1-y}F_yBiS_2$ for the first time. Figure 5(a), (c) and (e) show X-ray diffraction patterns for $Nd_{1-z}Sm_zO_{1-y}F_yBiS_2$ with $y$ = 0.7, 0.5 and 0.3, respectively. Almost all of the observed peaks are explained using the tetragonal *P*4/*nmm* space group. In the $Nd_{1-z}Sm_zO_{0.3}F_{0.7}BiS_2$ and $Nd_{1-z}Sm_zO_{0.5}F_{0.5}BiS_2$ samples, impurity phases are clearly observed when the Sm concentration $z$ is 0.8 as indicated by symbols of *, which indicates that there is a solubility limit of Sm with Nd for the RE site. For $Nd_{1-z}Sm_zO_{0.7}F_{0.3}BiS_2$, impurity phases are not observed for $z$ = 0 – 0.8. Figure 5(b), (d) and (f) show the enlarged X-ray diffraction patterns near the (200) peak for the $Nd_{1-z}Sm_zO_{1-y}F_yBiS_2$ samples with $y$ = 0.7, 0.5 and 0.3, respectively. As shown in Fig. 5(b) and (d), the peak position of the (200) peak systematically shifts to higher angles with increasing Sm concentration ($z$) in $Nd_{1-z}Sm_zO_{0.3}F_{0.7}BiS_2$ and $Nd_{1-z}Sm_zO_{0.5}F_{0.5}BiS_2$. In $Nd_{1-z}Sm_zO_{0.7}F_{0.3}BiS_2$, the peak position of the (200) peak largely shifts to higher angles from $z$ = 0.2 to $z$ = 0.4, while the peak position of the (200) peak does not show a remarkable change within $z$ = 0.4 – 0.8. There might be a local structure transition at a boundary between $z$ = 0.2 and 0.4. To clarify the relationship between RE (RE = Nd and Sm) concentration and lattice constants, the lattice constants $a$ and $c$ were calculated using the peak positions of the (200) and (004) peaks.

The lattice constants of $a$ and $c$ for all the samples of $Nd_{1-z}Sm_zO_{1-y}F_yBiS_2$ ($y$ = 0.3, 0.5 and 0.7) are plotted in Fig. 6(a) and (b), respectively. With increasing Sm concentration ($z$), the lattice constant of $a$ monotonously decreases for $y$ = 0.5 and 0.7 while the lattice constant of $c$ does not show a remarkable dependence on Sm concentration ($z$). The lattice constant of the $a$ axis for $y$ = 0.3 shows a large drop at around $z$ = 0.2 – 0.4. These results indicate that the partial substitution for Nd by Sm generates lattice contraction of $a$ axis. The lattice constants of $a$ in $Nd_{1-z}Sm_zO_{1-y}F_yBiS_2$ is shorter than that of the shortest value of $a$ axis in $Ce_{1-x}Nd_xO_{1-y}F_yBiS_2$. This indicates that the effect of chemical pressure in $Nd_{1-z}Sm_zO_{1-y}F_yBiS_2$ is clearly larger than that in the $Ce_{1-x}Nd_xO_{1-y}F_yBiS_2$ system.

Figure 7(a) shows the temperature dependence of magnetic susceptibility for $Nd_{1-z}Sm_zO_{0.3}F_{0.7}BiS_2$. In $Nd_{1-z}Sm_zO_{0.3}F_{0.7}BiS_2$, a superconducting transition is observed within $z \leq 0.6$ but $\Delta\chi$ is very small as observed in $Ce_{1-x}Nd_xO_{0.3}F_{0.7}BiS_2$. Namely, bulk superconductivity cannot be induced for $y$ = 0.7 by chemical pressure in a line of $Ce_{1-x}Nd_xO_{1-y}F_yBiS_2$ and $Nd_{1-z}Sm_zO_{1-y}F_yBiS_2$. Figure 7(b) shows the enlarged temperature dependence of magnetic susceptibility for $Nd_{1-z}Sm_zO_{0.3}F_{0.7}BiS_2$. The $T_c^{onset}$ decreases with increasing Sm concentration.

In $Nd_{1-z}Sm_zO_{0.5}F_{0.5}BiS_2$, superconductivity is observed within $0 \leq z \leq 0.6$. The $\Delta\chi$ is clearly larger than those of $Nd_{1-z}Sm_zO_{0.3}F_{0.7}BiS_2$, indicating the superconducting state is bulk in nature. But, the value of $\Delta\chi$ relatively decreases with increasing Sm concentration as shown in Fig. 7(c). The $T_c^{onset}$ increases with increasing Sm concentration as shown in Fig. 7(d) and reaches 5.5 K for $z$ = 0.6.

In $Nd_{1-z}Sm_zO_{0.7}F_{0.3}BiS_2$, bulk superconductivity is observed within $0 \leq z \leq 0.8$. Similarly to $Nd_{1-z}Sm_zO_{0.5}F_{0.5}BiS_2$,the $\Delta\chi$ tends to relatively decrease with increasing Sm concentration as shown in Fig. 7(e) but $T_c^{onset}$ increases with increasing Sm concentration as shown in Fig. 7(f) and reaches 5.6 K for $z$ = 0.8.

Figure 8 shows the Sm concentration dependence of $T_c^{onset}$ for $Nd_{1-z}Sm_zO_{1-y}F_yBiS_2$ ($y$ = 0.3, 0.5 and 0.7). The symbols of × indicate that the sample at this point does not exhibit a superconducting transition at above 2 K. Bulk superconductivity is observed for $0 \leq z \leq 0.6$ with $y$ = 0.5 and $0 \leq z \leq 0.8$ with $y$ = 0.3. The $T_c^{onset}$ increases with increasing Sm concentration for $y$ = 0.3 and 0.5 while the $T_c^{onset}$ decreases with increasing Sm concentration for $y$ = 0.7.

## 5. Phase diagram

To understand the obtained results and discuss relationship between the chemical pressure effect and the inducement of bulk superconductivity, we have established three kinds of phase diagram.

Figure 9 shows a chemical pressure-$\Delta\chi$ phase diagram for $Ce_{1-x}Nd_xO_{1-y}F_yBiS_2$ ($y$ = 0.3, 0.5 and 0.7) and $Nd_{1-z}Sm_zO_{1-y}F_yBiS_2$ ($y$ = 0.3, 0.5 and 0.7). The data points of $Ce_{1-x}Nd_xO_{1-y}F_yBiS_2$ are plotted in the left area of broken line. The data points of $Nd_{1-z}Sm_zO_{1-y}F_yBiS_2$ are shown in the right area of the broken line. The chemical pressure in the $Nd_{1-z}Sm_zO_{1-y}F_yBiS_2$ system is larger than that in the $Ce_{1-x}Nd_xO_{1-y}F_yBiS_2$ system because the ionic radius becomes smaller in the order of $Ce^{3+}$, $Nd^{3+}$ and $Sm^{3+}$. In this phase diagram, we can discuss whether the superconducting states induced by chemical pressure are bulk or filamentary. We note that the samples of $y$ = 0.7 do not exhibit bulk superconductivity because the values of $\Delta\chi$ for all the samples with $y$ = 0.7 are lower than 0.05 emu/Oe·cm$^3$. In fact, bulk superconductivity is not induced by chemical pressure for $y$ = 0.7. We consider that the absence of bulk superconductivity is due to the higher F concentration. The high F concentration may result in a crystal structure unfavorable for the appearance of superconductivity. One possibility is the formation of coupling between the REO blocking layer and the BiS superconducting plane. Recent x-ray absorption spectroscopy measurements indicated that the RE-Bi coupling via the out-plane S atom (RE-S-Bi channel) could form in $CeO_{1-y}F_yBiS_2$[21]. When the Ce-S-Bi channel is formed, both superconductivity and magnetic ordering are not observed in $CeO_{1-y}F_yBiS_2$. In contrast, both superconductivity and magnetic ordering are observed when the Ce-S-Bi channel is not formed. In $Ce_{1-x}Nd_xO_{0.3}F_{0.7}BiS_2$, however, magnetic ordering of the RE site is clearly observed for $0 \leq x \leq 0.8$. This indicates that there is no RE-S-Bi coupling. Therefore, we assume that the other factor affects the appearance of bulk superconductivity in the present $REO_{0.3}F_{0.7}BiS_2$ system. The possible reason is excess electron carriers within the $BiS_2$ layers, which may be unfavorable for the appearance of bulk superconductivity in the $REO_{0.3}F_{0.7}BiS_2$ ($y$ = 0.7) system. To clarify the reason for the absence of bulk superconductivity in $y$ = 0.7, studies using single crystals are needed.

For $y$ = 0.5 and 0.3, the value of $\Delta\chi$ clearly increases with increasing Nd concentration in $Ce_{1-x}Nd_xO_{1-y}F_yBiS_2$ up to $x$ = 1.0, indicating the inducement of bulk superconductivity by chemical pressure. Then, the value of $\Delta\chi$ slightly decreases with increasing Sm concentration in $Nd_{1-z}Sm_zO_{1-y}F_yBiS_2$, but those are still large value even with a higher Sm concentration ($z$). Here, we define a criterion of $\Delta\chi > 0.05$ emu/Oe·cm$^3$ for bulk superconducting states to discuss the essential relationship between chemical pressure effect, crystal structure and $T_c$.

To discuss the correlation between chemical pressure effect and $T_c$, we plotted a chemical pressure-$T_c$ phase diagram for the $Ce_{1-x}Nd_xO_{1-y}F_yBiS_2$ ($y$ = 0.3, 0.5 and 0.7) and $Nd_{1-z}Sm_zO_{1-y}F_yBiS_2$ ($y$ = 0.3, 0.5 and 0.7) in Fig. 10. The data points of $Ce_{1-x}Nd_xO_{1-y}F_yBiS_2$ are plotted in the left area of broken line. The data points of $Nd_{1-z}Sm_zO_{1-y}F_yBiS_2$ are shown in the right area of the broken line. The symbols of × indicate that the sample at this point does not exhibit a superconducting transition at above 2 K. The open symbols indicate the $T_c^{onset}$ in a filamentary superconductivity state. The filled symbols indicate the $T_c^{onset}$ in a bulk superconductivity state.

For the F concentration of $y$ = 0.7, filamentary superconductivity is observed for all the samples of $Ce_{1-x}Nd_xO_{1-y}F_yBiS_2$ and $Nd_{1-z}Sm_zO_{1-y}F_yBiS_2$ as shown in Fig. 10(a). On the superconducting transition temperature, the $T_c^{onset}$ increases with increasing chemical pressure in $Ce_{1-x}Nd_xO_{1-y}F_yBiS_2$, and it decreases with increasing chemical pressure in $Nd_{1-z}Sm_zO_{1-y}F_yBiS_2$.

For the F concentration of $y$ = 0.5, bulk superconductivity is observed for $x \geq 0.5$ as shown in Fig. 10(b). The $T_c^{onset}$ continues to increase with increasing chemical pressure in a wide range of $x$ = 0.4 – 1.0 and $z$ = 0 - 0.6. For $y$ = 0.3, bulk superconductivity is observed for $x$ = 1.0 in $Ce_{1-x}Nd_xO_{0.7}F_{0.3}BiS_2$ and all the samples of $Nd_{1-z}Sm_zO_{0.7}F_{0.3}BiS_2$ as shown in Fig. 10(c). The $T_c^{onset}$ increases with increasing chemical pressure in a wide range from $z$ = 0 ($x$ = 1.0) to $z$ = 0.8. Having considered the phase diagram for $y$ = 0.5 and 0.3, we find that the RE concentration, at which bulk superconductivity is induced, clearly depends on F concentration ($y$). In the $REO_{1-y}F_yBiS_2$ family, both lattice contraction of the *a* axis by chemical pressure and optimal F concentration seem to be essential for the inducement of bulk superconductivity.

Finally, we discuss the relationship between $T_c$ and lattice constants in detail. Figure 11 shows the *c*/*a* ratio dependence of $T_c^{onset}$ for $Ce_{1-x}Nd_xO_{1-y}F_yBiS_2$ ($y$ = 0.3 and 0.5) and $Nd_{1-z}Sm_zO_{1-y}F_yBiS_2$ ($y$ = 0.3 and 0.5). With the criterion of $\Delta\chi > 0.05$ emu/Oe·cm$^3$, we excluded the data points of filamentary superconductors in this plot. We note that the *c*/*a* ratio approaches a characteristic area around 3.39 when the $T_c$ becomes higher for both $y$ = 0.5 and 0.3 as indicated by an orange square in Fig. 11. This implies that a larger *c*/*a* ratio is important for a higher $T_c$ in the $Ce_{1-x}Nd_xO_{1-y}F_yBiS_2$ and $Nd_{1-z}Sm_zO_{1-y}F_yBiS_2$ system. The uniaxial lattice contraction (chemical pressure) along the *a* axis is the strongest at this region. Therefore, both uniaxial contraction along the *a* axis generated by chemical pressure and optimal F concentration are essential for the inducement of bulk superconductivity with a higher $T_c$ in the $REO_{1-y}F_yBiS_2$ superconductors.

**6. Conclusion**

We have systematically investigated the crystal structure and magnetic properties of $BiS_2$-based superconductors $Ce_{1-x}Nd_xO_{1-y}F_yBiS_2$ ($y$ = 0.3, 0.5 and 0.7) and $Nd_{1-z}Sm_zO_{1-y}F_yBiS_2$ ($y$ = 0.3, 0.5 and 0.7) to investigate the chemical pressure effects on superconductivity of the $REO_{1-y}F_yBiS_2$ systems. In the $Ce_{1-x}Nd_xO_{1-y}F_yBiS_2$ samples, bulk superconductivity is observed for $x \geq 0.5$ with $y = 0.5$ and $x = 1.0$ with $y = 0.3$. The $Ce_{1-x}Nd_xO_{0.3}F_{0.7}BiS_2$ samples exhibit filamentary superconductivity. The $T_c^{onset}$ in $Ce_{1-x}Nd_xO_{1-y}F_yBiS_2$ commonly increases with increasing Nd concentration. The highest $T_c^{onset}$ in $Ce_{1-x}Nd_xO_{1-y}F_yBiS_2$ system is 4.8 K for $x = 1.0$ with $y = 0.5$. X-ray diffraction analysis indicates that the systematic substitution of Ce by Nd generates chemical pressure along the *a* axis while *c* axis does not show a remarkable change on Nd concentration. The lattice constants of *c* get shorter with increasing F concentration. This result indicates the partial substitution of O by F generates lattice contraction along *c* axis.

Furthermore, we found that Nd can be replaced by smaller Sm ions, up to $z = 0.8$. Bulk superconductivity is observed within $0 \leq z \leq 0.8$ with $y = 0.3$ and $0 \leq z \leq 0.6$ with $y = 0.5$. The $T_c^{onset}$ in $Nd_{1-z}Sm_zO_{1-y}F_yBiS_2$ increased with increasing Sm concentration. The highest $T_c$ is 5.6 K for $z = 0.8$ with $y = 0.3$. The $Nd_{1-z}Sm_zO_{0.3}F_{0.7}BiS_2$ samples exhibit filamentary superconductivity within a range of $0 \leq z \leq 0.4$. With increasing Sm concentration, the lattice constant of *a* decreases while the lattice constant of *c* does not show a remarkable dependence on Sm concentration. The lattice constant of *a* in $Nd_{1-z}Sm_zO_{1-y}F_yBiS_2$ is shorter than that of the shortest value of the lattice constant of *a* in $Ce_{1-x}Nd_xO_{1-y}F_yBiS_2$. This indicates that the effect of chemical pressure in $Nd_{1-z}Sm_zO_{1-y}F_yBiS_2$ is clearly larger than that in the $Ce_{1-x}Nd_xO_{1-y}F_yBiS_2$ system.

We established a three kinds of phase diagram for the $Ce_{1-x}Nd_xO_{1-y}F_yBiS_2$ ($y$ = 0.3, 0.5 and 0.7) and $Nd_{1-z}Sm_zO_{1-y}F_yBiS_2$ ($y$ = 0.3, 0.5 and 0.7). On the basis of the phase diagrams, both lattice contraction of *a* axis and optimal F concentration are essential for the inducement of bulk superconductivity in $REO_{1-y}F_yBiS_2$. Furthermore, the $c/a$ ratio of the higher $T_c$ samples is concentrated at near $c/a$ = 3.39. Within this region, the chemical pressure along the *a* axis is the strongest in the studied samples. Therefore, both uniaxial contraction along the *a* axis generated by chemical pressure and optimal F concentration are important for the inducement of bulk superconductivity with a higher $T_c$ in the $REO_{1-y}F_yBiS_2$ superconductors.

**Acknowledgements**

This work was partly supported by JSPS KAKENHI Grant Numbers 25707031, 26600077.

References

[1] J. G. Bednorz and K.A. Muiller, Z. phys. B **64,** 189 (1896).

[2] Y. Kamihara, T. Watanabe, M. Hirano, and H. Hosono, J. Am. Chem. Soc. **130**, 3296 (2008).

[3] Y. Mizuguchi, H. Fujihisa, Y. Gotoh, K. Suzuki, H. Usui, K. Kuroki, S. Demura, Y. Takano, H. Izawa, and O. Miura, Phys. Rev. B. **86,** 220510 (2012).

[4] Y. Mizuguchi, S. Demura, K. Deguchi, Y. Takano, H. Fujihisa, Y. Gotoh, H. Izawa, and O. Miura, J. Phys. Soc. Jpn. **81,** 114725 (2012).

[5] J. Xing, S. Li, X. Ding, H. Yang, H.H. Wen, Phys. Rev. B. **86,** 214518 **(**2012).

[6] S. Demura, K. Deguchi, Y. Mizuguchi, K. Sato, R. Honjyo, A. Yamashita, T. Yamaki, H. Hara, T. Watanabe, S. J. Denholme, M. Fujioka, H. Okazaki, T. Ozaki, O. Miura, T. Yamaguchi, H. Takeya, and Y. Takano, arXiv:1311.4267 (2013).

[7] R. Jha, A. Kumar, S. K. Singh, and V. P. S. Awana, J. Sup. Novel Mag. **26,** 499 (2013).

[8] J. Kajitani, K. Deguchi, T. Hiroi, A. Omachi, S. Demura, Y. Takano, O. Miura, and Y. Mizuguchi, J. Phys. Soc. Jpn. **83**, 065002 (2014).

[9] S. Demura, Y. Mizuguchi, K. Deguchi, H. Okazaki, H. Hara, T. Watanabe, S. J. Denholme, M. Fujioka, T. Ozaki, H. Fujihisa, Y. Gotoh, O. Miura, T. Yamaguchi, H. Takeya, and Y. Takano, J. Phys. Soc. Jpn. **82,** 033708 (2013).

[10] X. Lin, X. Ni, B. Chen, X. Xu, X. Yang, J. Dai, Y. Li, X. Yang, Y. Luo, Q. Tao, G. Cao, and Z. Xu, Phys. Rev. B **87,** 020504 (2013).

[11] H. Sakai, D. Kotajima, K. Saito, H. Wadati, Y. Wakisaka, M. Mizumaki, K. Nitta, Y. Tokura, and S. Ishiwata, J. Phys. Soc. Jpn. **83,** 014709 (2014).

[12] H. Usui, K. Suzuki, and K. Kuroki, Phys. Rev. B **86**, 220501 (2012).

[13] H. Kotegawa, Y. Tomita, H. Tou, H. Izawa, Y. Mizuguchi, O. Miura, S. Demura, K. Deguchi, and Y. Takano, J. Phys. Soc. Jpn. **81**, 103702 (2012).

[14] C. T. Wolowiec, D. Yazici, B. D. White, K. Huang, and M. B. Maple, Phys. Rev. B **88**, 064503 (2013).

[15] C. T. Wolowiec, B. D. White, I. Jeon, D. Yazici, K. Huang, and M. B. Maple: J. Phys.: Condens. Matter **25,** 422201 (2013).

[16] T. Tomita, M. Ebata, H. Soeda, H. Takahashi, H. Fujihisa, Y. Gotoh, Y. Mizuguchi, H. Izawa, O. Miura, S. Demura, K. Deguchi, and Y. Takano, J. Phys. Soc. Jpn. **83**, 063704 (2014).

[17] K. Deguchi, Y. Mizuguchi, S. Demura, H. Hara, T. Watanabe, S. J. Denholme, M. Fujioka, H. Okazaki, T. Ozaki, H. Takeya, T. Yamaguchi, O. Miura, and Y. Takano, EPL **101**, 17004 (2013).

[18] J. Kajitani, K. Deguchi, A. Omachi, T. Hiroi, Y. Takano, H. Takatsu, H. Kadowaki, O. Miura, and Y. Mizuguchi, Solid State Commun. **181**, 1 (2014).

[19] Y. Mizuguchi, T. Hiroi, J. Kajitani, H. Takatsu, H. Kadowaki, and O. Miura, J. Phys. Soc. Jpn. **83**, 053704 (2014).

[20] J. Kajitani, Atsushi Omachi, T. Hiroi, O. Miura, and Y. Mizuguchi, Physica C **504**, 33-35 (2014).

[21] T. Sugimoto, B. Joseph, E. Paris, A. Iadecola, T. Mizokawa, S. Demura, Y. Mizuguchi, Y. Takano, and N. L. Saini, Phys. Rev. B. **89**, 201117(R) (2014).

Figure captions

Fig. 1. (a) X-ray diffraction patterns for $Ce_{1-x}Nd_xO_{0.3}F_{0.7}BiS_2$. (b) Enlarged X-ray diffraction patterns near the (200) peak for $Ce_{1-x}Nd_xO_{0.3}F_{0.7}BiS_2$. (c) X-ray diffraction patterns for $Ce_{1-x}Nd_xO_{0.5}F_{0.5}BiS_2$. (d) Enlarged X-ray diffraction patterns near the (200) peak for $Ce_{1-x}Nd_xO_{0.5}F_{0.5}BiS_2$. (e) X-ray diffraction patterns for $Ce_{1-x}Nd_xO_{0.7}F_{0.3}BiS_2$. (f) Enlarged X-ray diffraction patterns near the (200) peak for $Ce_{1-x}Nd_xO_{0.7}F_{0.3}BiS_2$.

Fig. 2. Nd concentration dependence of the lattice constants of (a) $a$ and (b) $c$ for $Ce_{1-x}Nd_xO_{1-y}F_yBiS_2$ ($y$ = 0.3, 0.5 and 0.7).

Fig. 3. (a) Temperature dependence of magnetic susceptibility for $Ce_{1-x}Nd_xO_{0.3}F_{0.7}BiS_2$ ($x$ = 0, 0.2 and 0.4). (b) Temperature dependence of magnetic susceptibility for $Ce_{1-x}Nd_xO_{0.3}F_{0.7}BiS_2$ ($x$ = 0.6, 0.8 and 1.0). (c) Temperature dependence of magnetic susceptibility for $Ce_{1-x}Nd_xO_{0.5}F_{0.5}BiS_2$.

Fig. 4 Nd concentration dependence of $T_c^{onset}$ for $Ce_{1-x}Nd_xO_{1-y}F_yBiS_2$ ($y$ = 0.3, 0.5 and 0.7). The symbols of × indicate that the sample at this point does not exhibit a superconducting transition at above 2 K.

Fig. 5 (a) X-ray diffraction patterns for $Nd_{1-z}Sm_zO_{0.3}F_{0.7}BiS_2$. (b) Enlarged X-ray diffraction patterns near the (200) peak for $Nd_{1-z}Sm_zO_{0.3}F_{0.7}BiS_2$. (c) X-ray diffraction patterns for $Nd_{1-z}Sm_zO_{0.5}F_{0.5}BiS_2$. (d) Enlarged X-ray diffraction patterns near the (200) peak for $Nd_{1-z}Sm_zO_{0.5}F_{0.5}BiS_2$. (e) X-ray diffraction patterns for $Nd_{1-z}Sm_zO_{0.7}F_{0.3}BiS_2$. (f) Enlarged X-ray diffraction patterns near the (200) peak for $Nd_{1-z}Sm_zO_{0.7}F_{0.3}BiS_2$.

Fig. 6 Sm concentration dependence of the lattice constants of (a) $a$ and (b) $c$ for $Nd_{1-z}Sm_zO_{1-y}F_yBiS_2$ ($y$ = 0.3, 0.5 and 0.7).

Fig. 7. (a) Temperature dependence of magnetic susceptibility for $Nd_{0.3}Sm_{0.7}O_{1-y}F_yBiS_2$. (b) Enlarged temperature dependence of magnetic susceptibility for $Nd_{1-z}Sm_zO_{0.3}F_{0.7}BiS_2$. (c) Temperature dependence of magnetic susceptibility for $Nd_{1-z}Sm_zO_{0.5}F_{0.5}BiS_2$. (d) Enlarged temperature dependence of magnetic susceptibility for $Nd_{1-z}Sm_zO_{0.5}F_{0.5}BiS_2$. (e) Temperature dependence of magnetic susceptibility for $Nd_{1-z}Sm_zO_{0.7}F_{0.3}BiS_2$. (f) Enlarged temperature dependence of magnetic susceptibility for $Nd_{1-z}Sm_zO_{0.7}F_{0.3}BiS_2$.

Fig. 8. Sm concentration dependence of $T_c^{onset}$ for $Nd_{1-z}Sm_zO_{1-y}F_yBiS_2$ ($y$ = 0.3, 0.5 and 0.7). The symbols of × indicate that the sample at this point does not exhibit a superconducting transition at above 2 K.

Fig. 9. Chemical pressure-$\Delta\chi$ phase diagram for $Ce_{1-x}Nd_xO_{1-y}F_yBiS_2$ ($y$ = 0.3, 0.5 and 0.7) and $Nd_{1-z}Sm_zO_{1-y}F_yBiS_2$ ($y$ = 0.3, 0.5 and 0.7).

Fig. 10. Chemical pressure-$T_c$ phase diagram for $Ce_{1-x}Nd_xO_{1-y}F_yBiS_2$ and $Nd_{1-z}Sm_zO_{1-y}F_yBiS_2$ with (a) $y$ = 0.3, (b) $y$ = 0.5 and (c) $y$ = 0.7. The symbols of × indicate that the sample at this point does not exhibit a superconducting transition at above 2 K.

Fig. 11. $c/a$ dependence of $T_c^{onset}$ for $Ce_{1-x}Nd_xO_{1-y}F_yBiS_2$ ($y$ = 0.3 and 0.5) and $Nd_{1-z}Sm_zO_{1-y}F_yBiS_2$ ($y$ = 0.3 and 0.5). A higher $T_c$ is obtained at around the area indicated with an orange square.

Figure 1

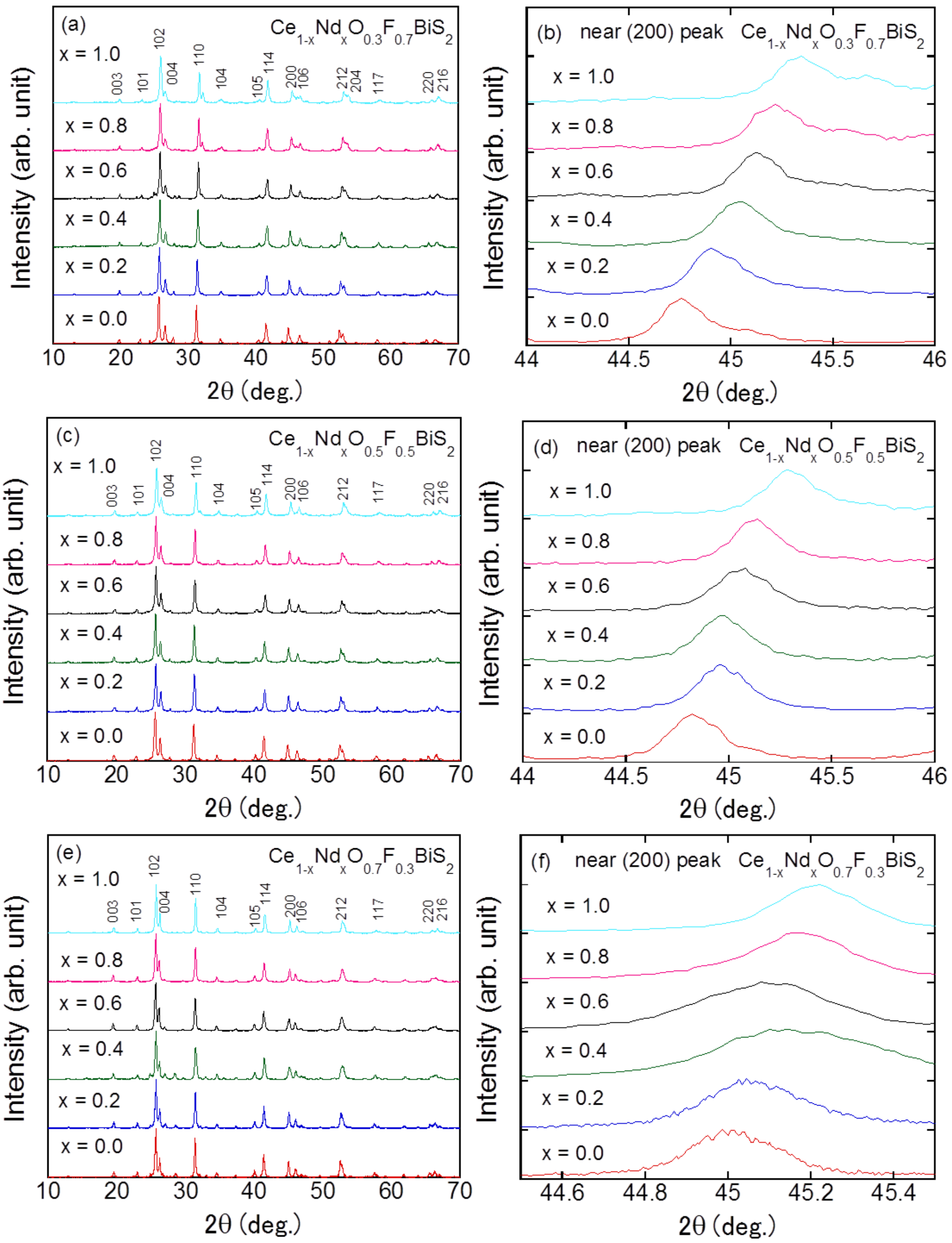

Figure 2

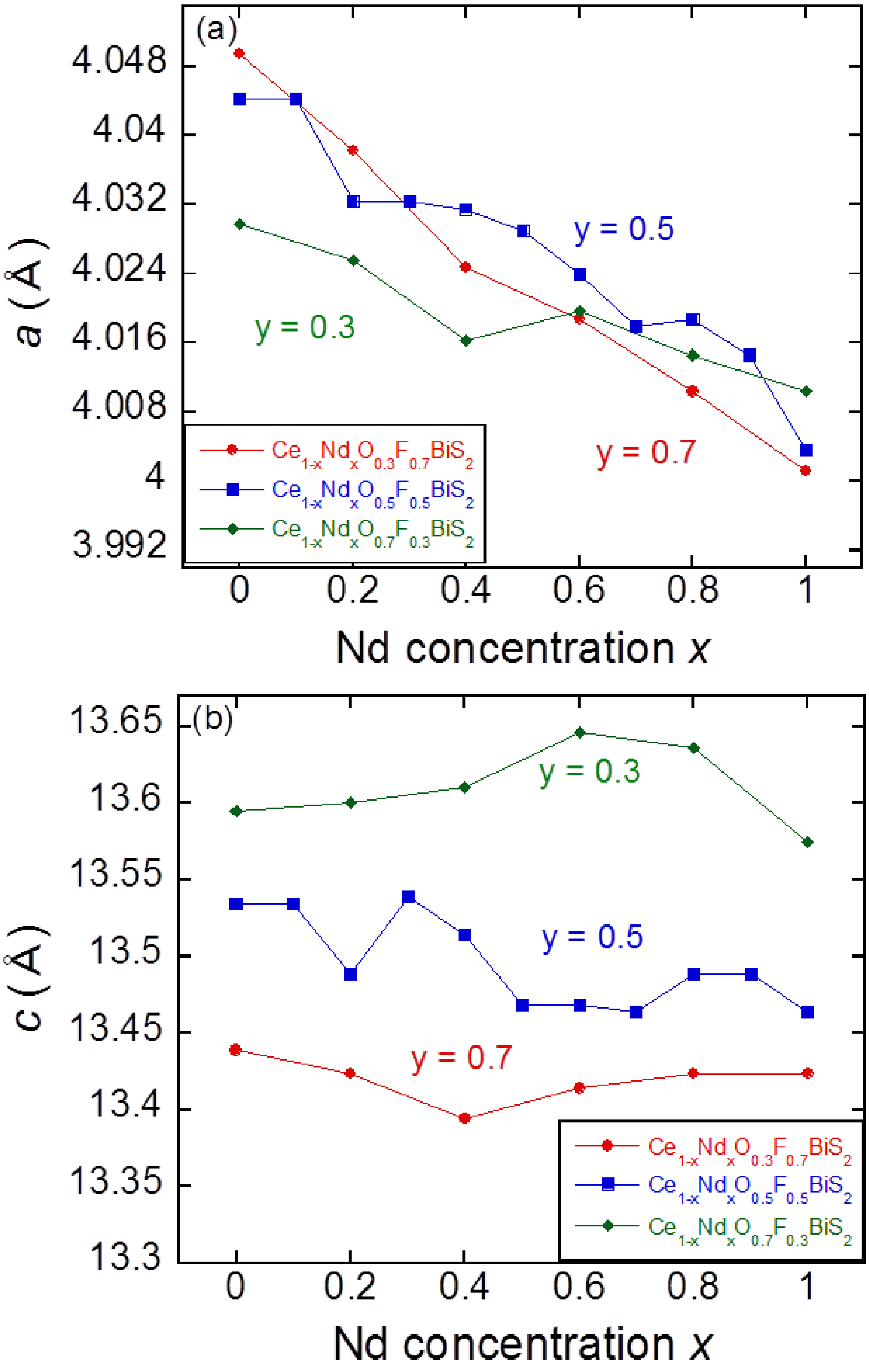

Figure 3

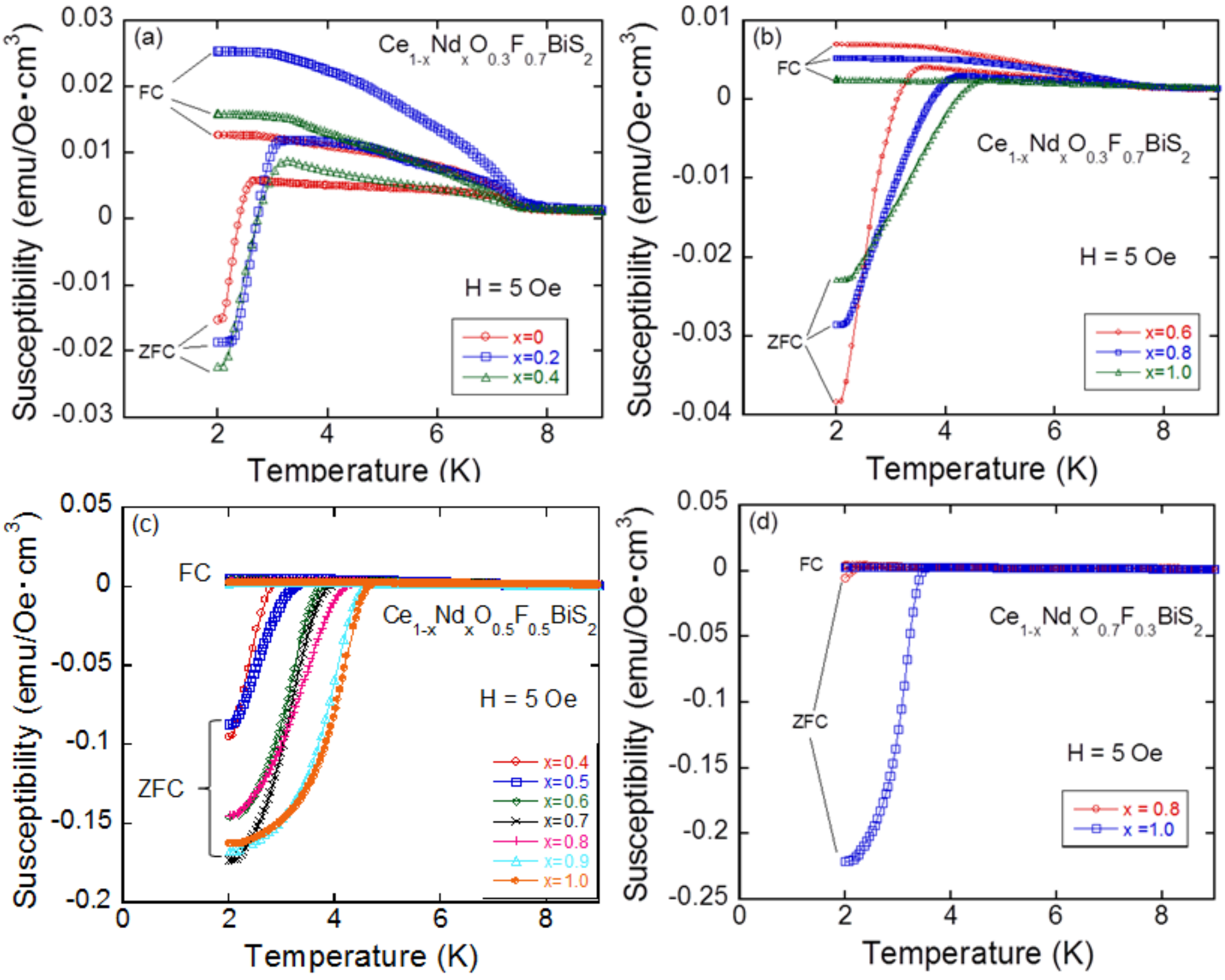

Figure 4

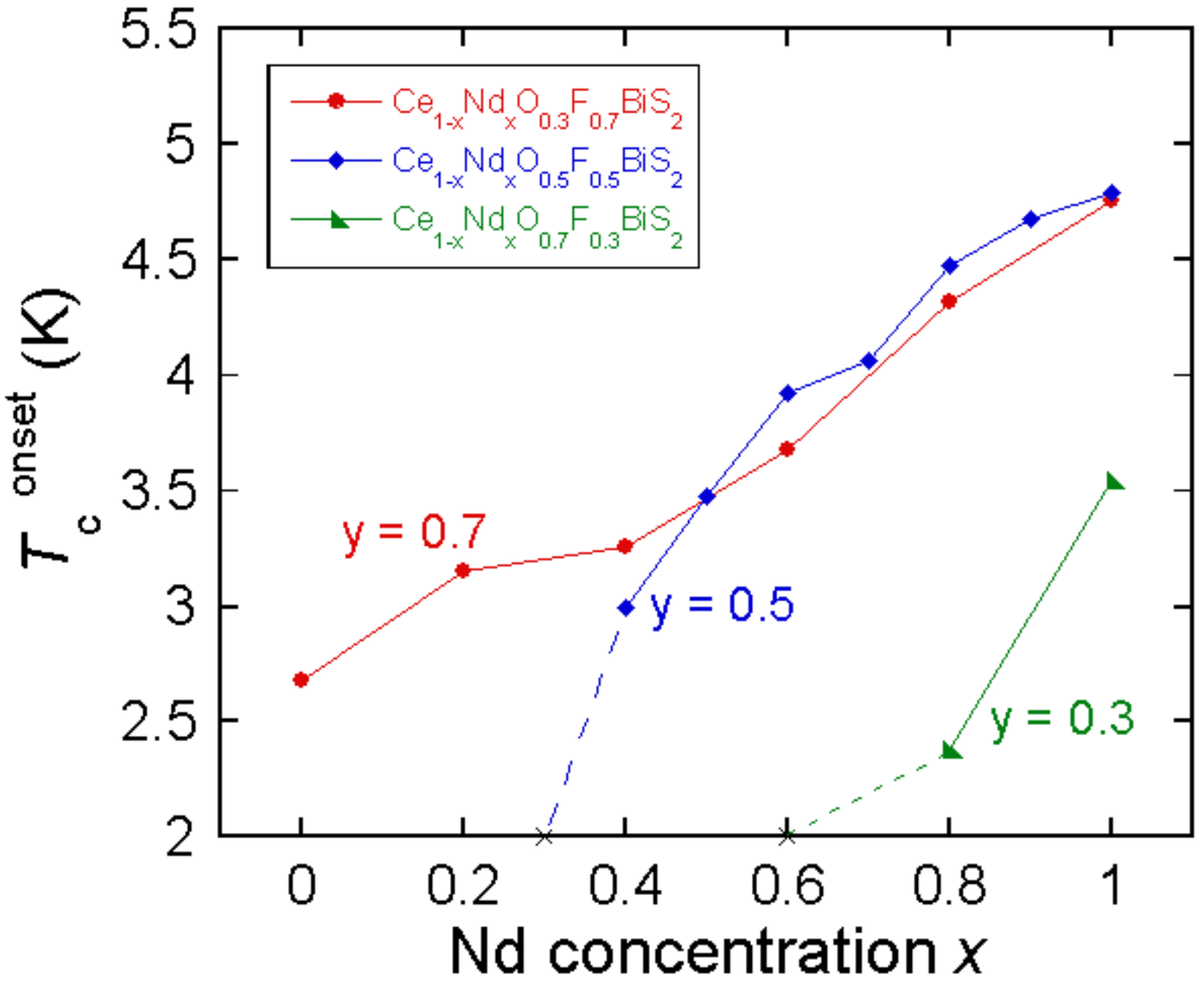

Figure 5

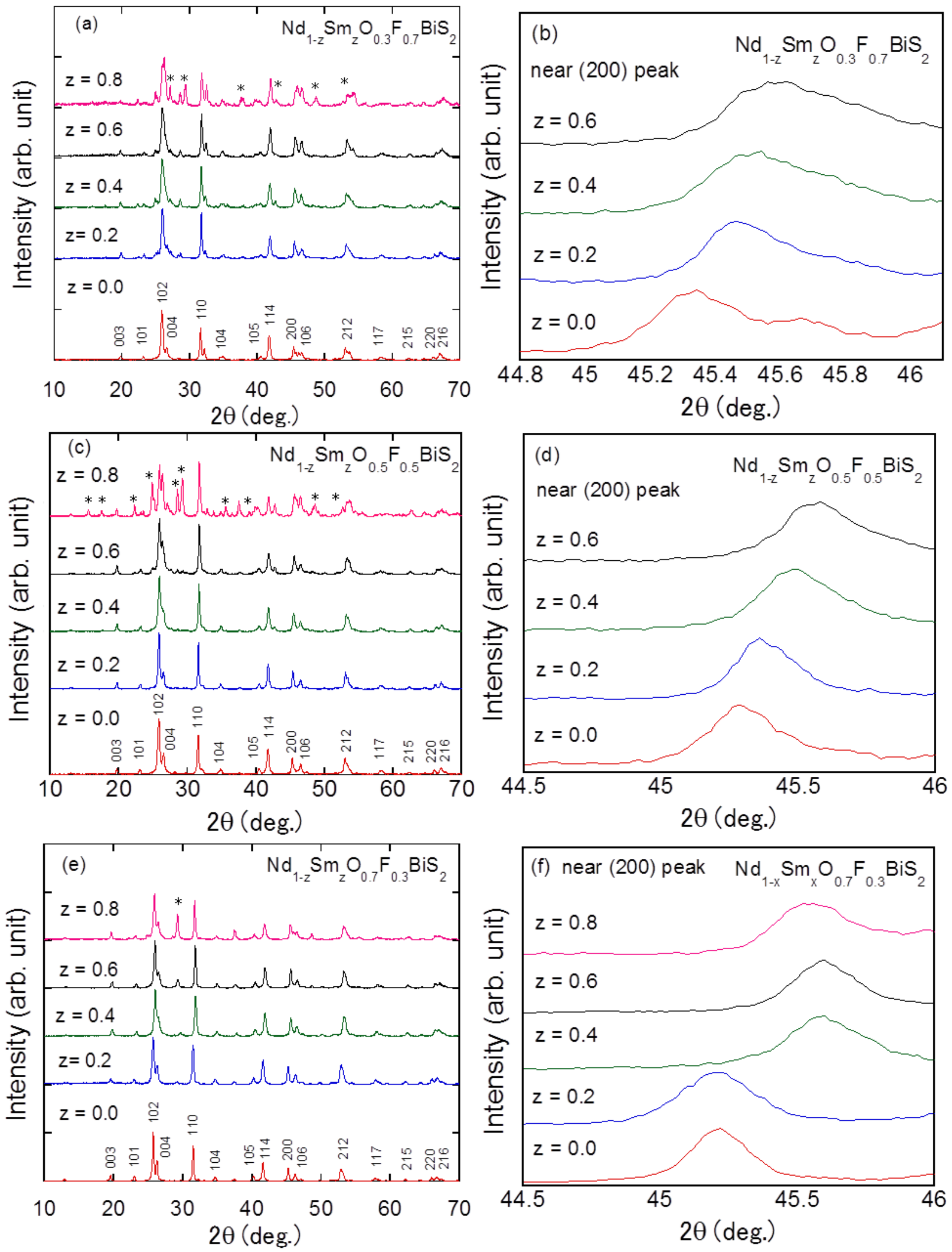

Figure 6

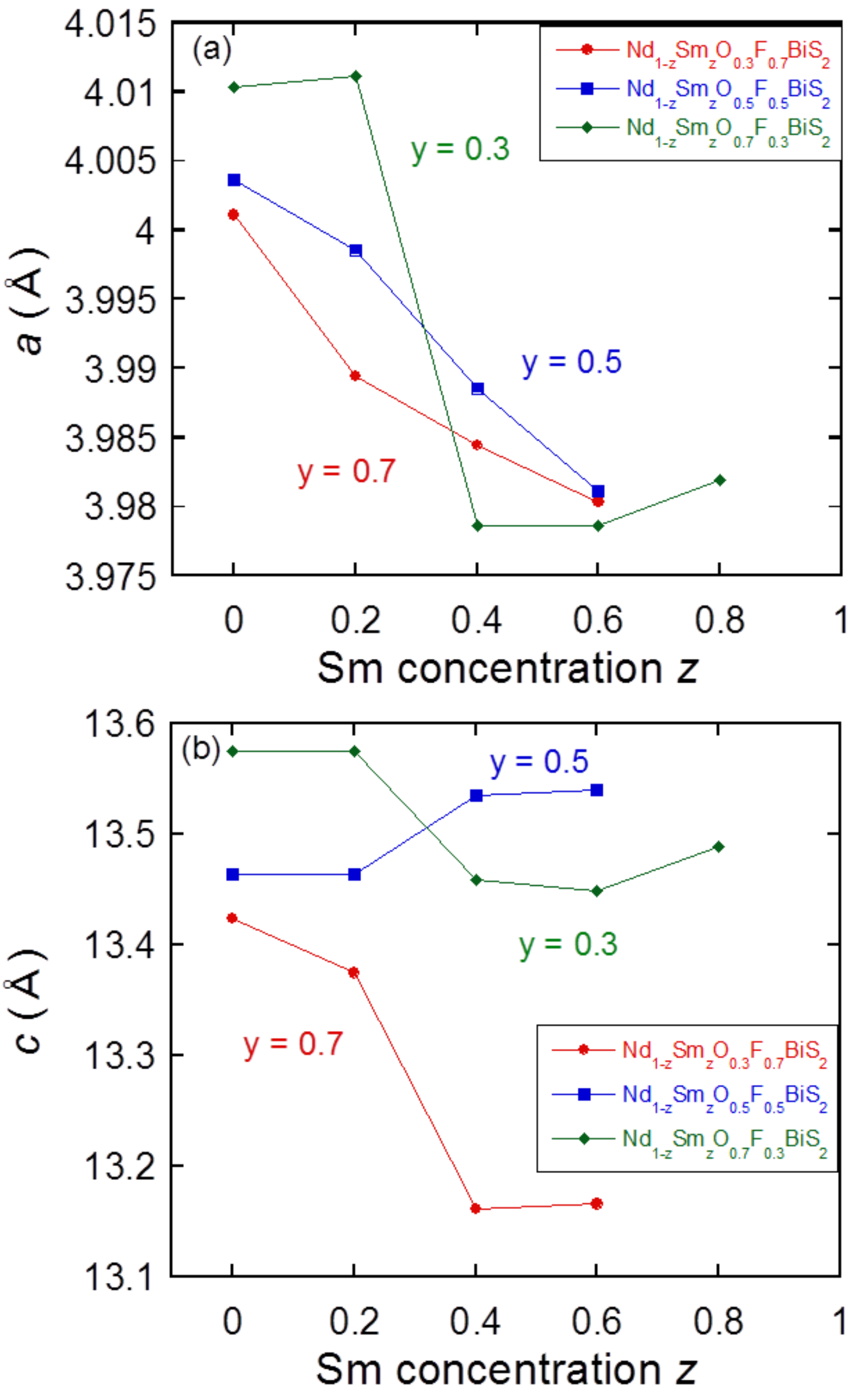

Figure 7

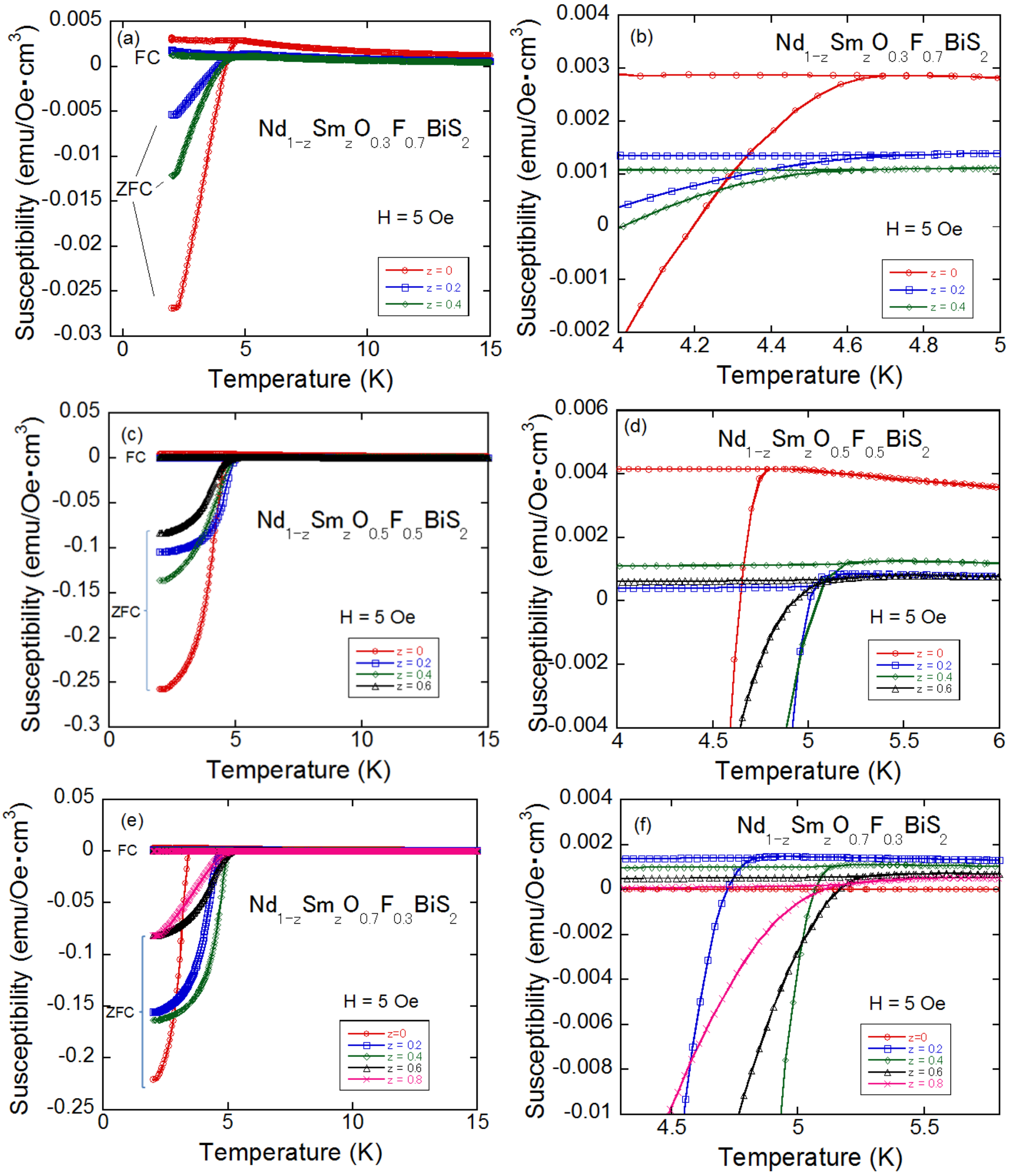

Figure 8

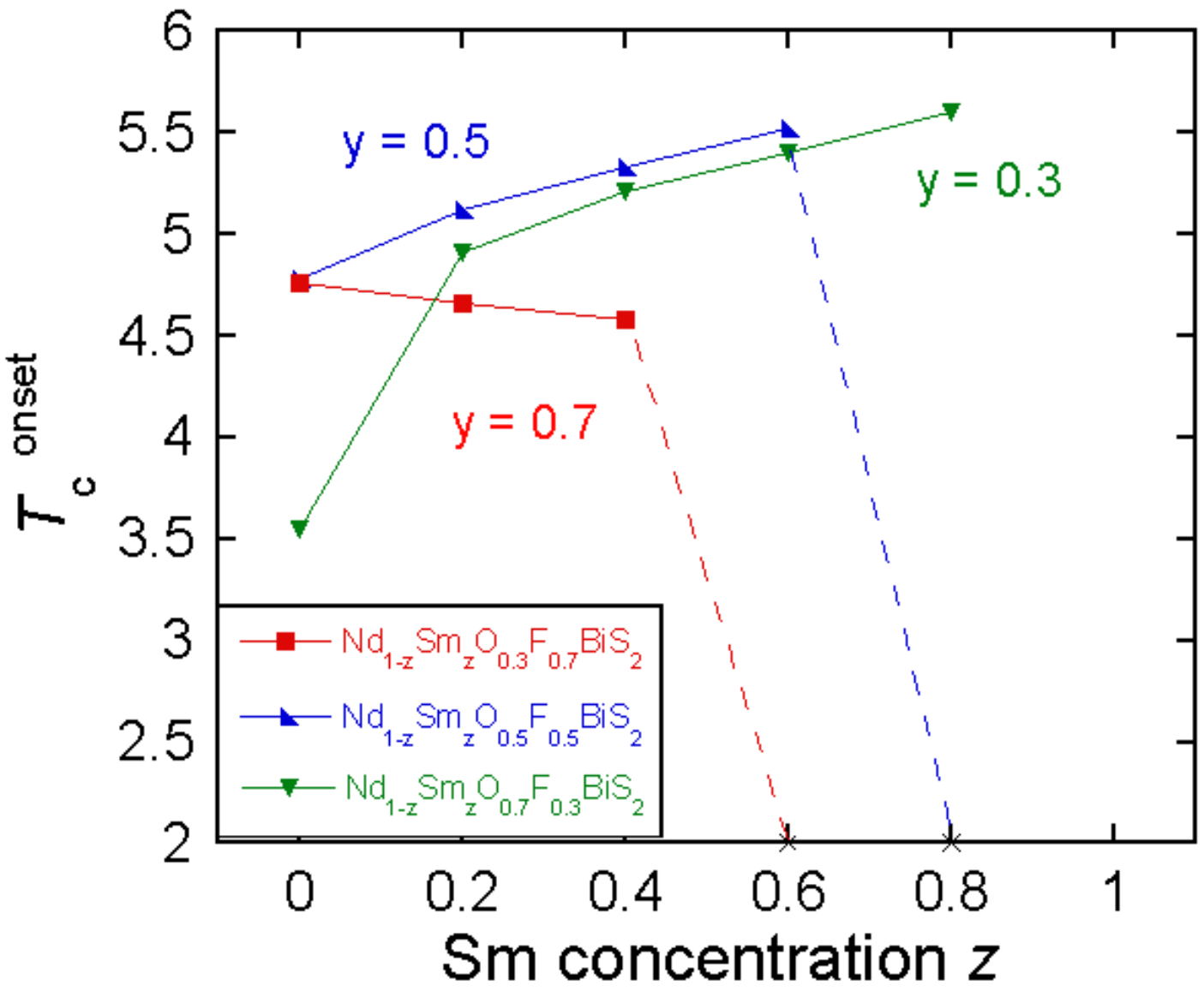

Figure 9

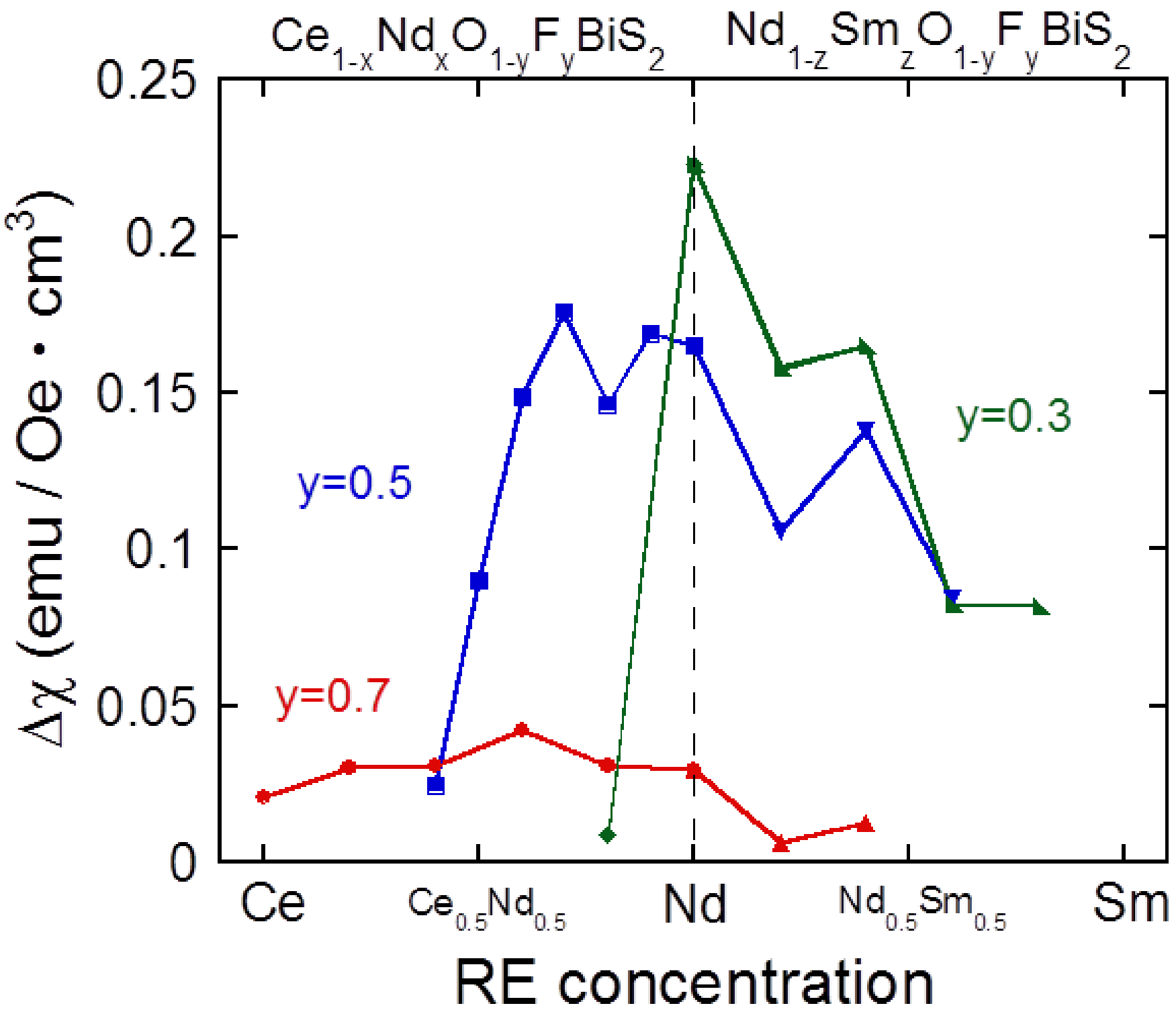

Figure 10

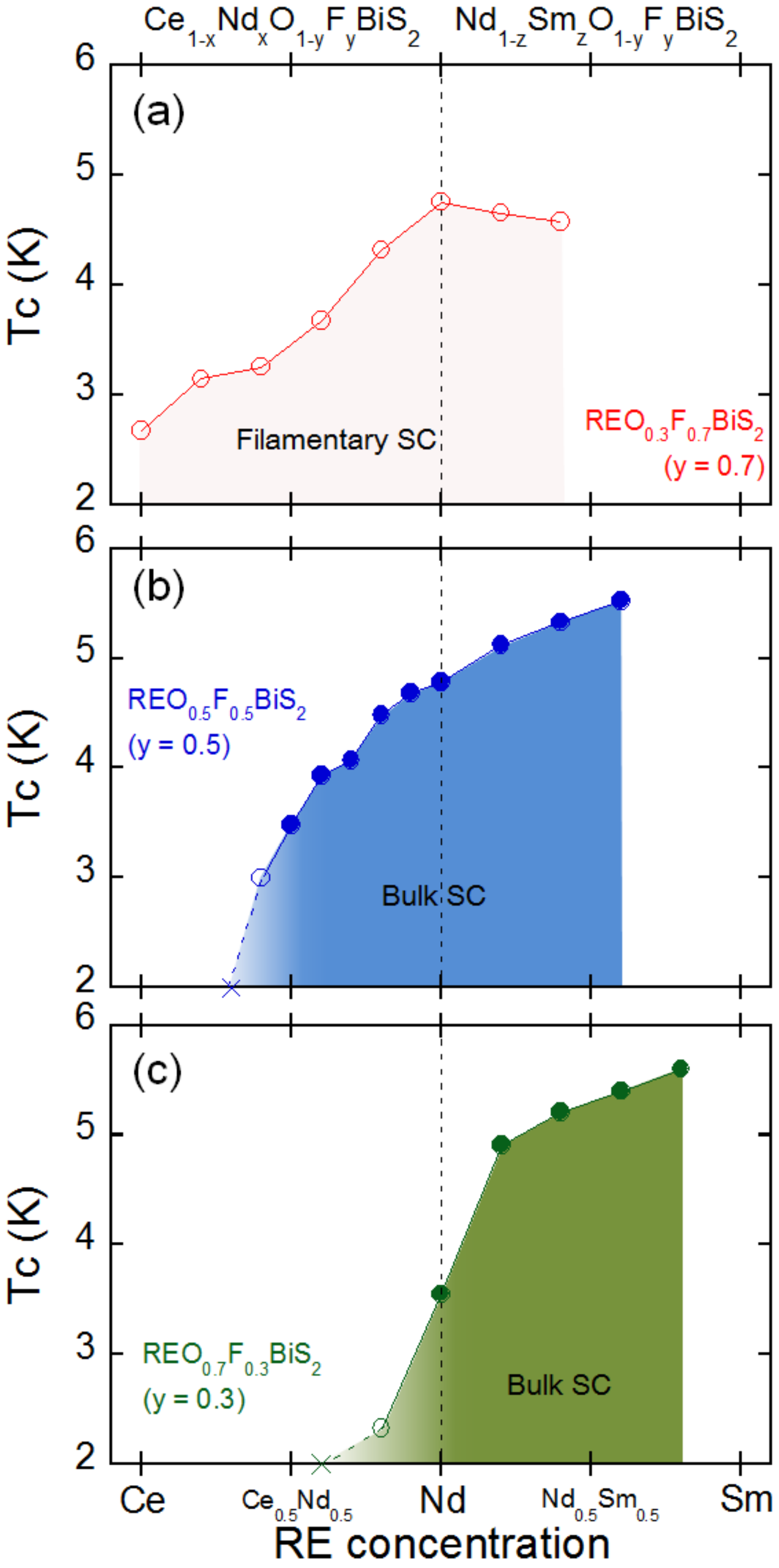

Figure 11

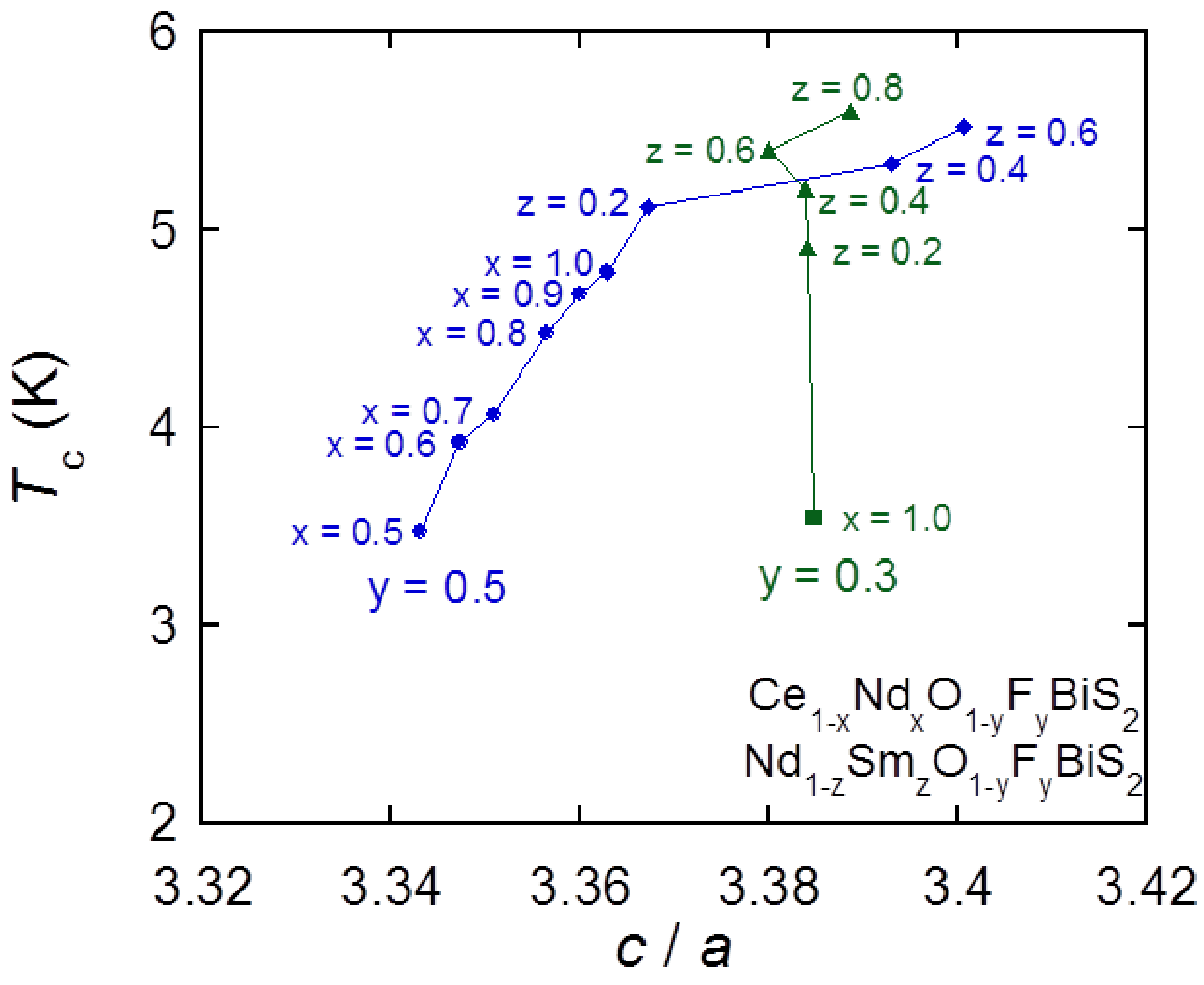